\documentclass[a4paper,fleqn,usenatbib]{mnras}

\usepackage[T1]{fontenc}
\usepackage{ae,aecompl}

\usepackage{times, graphicx, setspace, latexsym, amssymb, amsmath, fancyhdr}
\usepackage{longtable}

\def\lsim{\mathrel{\rlap{\lower4pt\hbox{\hskip1pt$\sim$}}
    \raise1pt\hbox{$<$}}}                
\def\gsim{\mathrel{\rlap{\lower4pt\hbox{\hskip1pt$\sim$}}
    \raise1pt\hbox{$>$}}}                

\title[HTRU XV: timing of 25 PSRs]{The High Time Resolution Universe
  Pulsar Survey -- XV:\\completion of the intermediate latitude survey
  with the discovery and timing of 25 further pulsars}

\author[M. Burgay et al.]{M.
  Burgay$^{1}$\thanks{E-mail:marta.burgay@inaf.it},
B. Stappers$^{2}$,
M.~Bailes,$^{3,4}$
E.~D.~Barr,$^{5,4}$
S.~Bates,$^{2}$
N.~D.~R.~Bhat,$^{6}$
\newauthor
S.~Burke-Spolaor,$^{7}$
A.~D.~Cameron,$^{8}$
D.~J.~Champion,$^{5}$ 
R.~P.~Eatough,$^{5}$
C.~M.~L.~Flynn,$^{3}$
\newauthor
A.~Jameson,$^{3,4}$
S.~Johnston,$^{8}$
M.~J.~Keith,$^{2}$
E.~F.~Keane,$^{9}$
M.~Kramer,$^{5,2}$
L.~Levin,$^{2}$
C.~Ng,$^{10}$
\newauthor
E.~Petroff,$^{11}$
A.~Possenti,$^{1}$
W.~van Straten,$^{12}$
C.~Tiburzi,$^{5,13}$
L.~Bondonneau,$^{14}$
A.G.~Lyne$^{2}$
\\
$^{1}$INAF - Osservatorio Astronomico di Cagliari, via della Scienza 5,
09047 Selargius (CA), Italy\\
$^{2}$Jodrell Bank Centre for Astrophysics, University of Manchester, Alan Turing Building, Oxford Road, Manchester M13 9PL, United Kingdom\\
$^{3}$Centre for Astrophysics and Supercomputing, Swinburne University of Technology, Mail H39, PO Box 218, VIC 3122, Australia. \\
$^{4}$ARC Centre of Excellence for Gravitational Wave Discovery (OzGrav), Swinburne University of Technology, Mail H11, PO Box 218, VIC 3122, Australia.\\
$^{5}$Max-Planck-Institut f\"{u}r Radioastronomie, Auf dem H\"{u}gel 69, D-53121 Bonn, Germany. \\
$^{6}$International Centre for Radio Astronomy Research, Curtin University, Bentley, WA 6102, Australia.\\
$^{7}$Center for Gravitational Waves and Cosmology, West Virginia University, Chestnut Ridge Research Building, Morgantown, WV 26505, USA.\\
$^{8}$CSIRO Astronomy \& Space Science, Australia Telescope National Facility, P.O. Box 76, Epping, NSW 1710, Australia.\\
$^{9}$SKA Organisation, Jodrell Bank Observatory, SK11 9DL, UK.\\
$^{10}$Department of Physics and Astronomy, University of British Columbia, 6224 Agricultural Road, Vancouver, BC V6T 1Z1, Canada.\\
$^{11}$ASTRON, the Netherlands Institute for Radio Astronomy, Postbus 2, NL-7990 AA Dwingeloo, the Netherlands.\\
$^{12}$Institute for Radio Astronomy \& Space Research, Auckland University of Technology, Private Bag 92006, Auckland 1142, New Zealand.\\
$^{13}$Fakult\"{a}t fur Physik, Universit\"{a}t Bielefeld, Postfach 100131, D-33501 Bielefeld, Germany.\\
$^{14}$LPC2E - Universit\'e d'Orl\'eans / CNRS, F-45071 Orl\'eans CEDEX 2, France.\\
}

\date{Accepted XXX. Received YYY; in original form ZZZ}

\pubyear{2018}

\begin{document}
\label{firstpage}
\pagerange{\pageref{firstpage}--\pageref{lastpage}}
\maketitle

\begin{abstract}
We report on the latest six pulsars discovered through our standard
pipeline in the intermediate-latitude region ($|b|$ $<$ 15$^{\circ}$)
of the Parkes High Time Resolution Universe Survey (HTRU). We also
present timing solutions for the new discoveries and for 19 further
pulsars for which only discovery parameters were previously
published. Highlights of the presented sample include the isolated
millisecond pulsar J1826$-$2415, the long-period binary pulsar
J1837$-$0822 in a mildly eccentric 98-day orbit with a $>$ 0.27
M$_{\odot}$ companion, and the nulling pulsar J1638$-$4233, detected
only 10\% of the time. Other interesting objects are PSR J1757$-$1500,
exhibiting sporadic mode changes, and PSR J1635$-$2616 showing one
glitch over 6 years. The new discoveries bring the total count of HTRU
intermediate-latitude pulsars to 113, 25\% of which are recycled
pulsars. This is the higest ratio of recycled over ordinary pulsars
discoveries of all recent pulsar surveys in this region of the
sky. Among HTRU recycled pulsars, four are isolated objects. Comparing
the characteristics of Galactic fully-recycled isolated MSPs with those
of eclipsing binaries (``spiders''), from which the former are believed
to have formed, we highlight a discrepancy in their spatial
distribution. This may reflect a difference in the natal kick, hence,
possibly, a different formation path. On the other hand, however,
isolated fully-recycled MSPs spin periods are, on average, longer than
those of spiders, in line with what one would expect, from simple
magnetic-dipole spin-down, if the former were indeed evolved from the
latter.


\end{abstract}




\section{Introduction}

Pulsars, and in particular the fastest rotating ones, are invaluable
tools to explore many aspects of astrophysics and physics: stellar
evolution, the interstellar medium and the Galactic magnetic field,
plasma physics, relativistic gravity -- including gravitational wave
emission -- and nuclear physics are some of the fields that pulsar
studies are contributing to \citep[see e.g.][and references
  therein]{ls90,lk05}. Finding more pulsars, and especially
millisecond pulsars (MSPs), has led to important steps forward in all
of the above topics, and more.

The High Time Resolution Universe Survey \citep[HTRU;][]{kjv+10} is an
experiment searching the majority of the sky visible from the Parkes
64-m radio telescope (NSW, Australia) for radio pulsars and fast
transients. The survey, started in 2008 and completed in 2014, is
divided into three parts: the first \citep{ncb+15} covers the lower
latitudes ($|b|$ $<$ 3.5$^{\circ}$) with 70-minute pointings, to
search for faint pulsars and relativistic binaries in the depths of
the Galactic plane; the second covers intermediate latitudes ($|b|$
$<$ 15$^{\circ}$) and a longitude range from $-$120$^{\circ}$ to
30$^{\circ}$, with 9-minute pointings, with the main aim of finding
relatively bright MSPs to be added to pulsar timing array experiments
\citep{bbb+12}; the third covers the rest of the sky accessible at
Parkes (with declination $<$ 10$^{\circ}$) with 4-minute long
pointings, mainly to search for fast transient events, such as Fast
Radio Bursts (FRBs; \citet{lbm+07,tsb+13}). The Parkes survey is
complemented by a twin experiment \citep[HTRU-North;][]{bck+13} carried
out at the Effelsberg 100-m radio telescope (Germany).

In this paper we report on the last discoveries made by our standard
pipeline \citep[described in][]{kjv+10} in the intermediate-latitude
part of the survey ({\it{HTRU-med}}, in the following), and on the
timing results of the follow-up campaigns on the new pulsars as well
as of 19 other HTRU-med sources for which only discovery parameters
have been previoulsy published \citep{bbb+12}.\footnote{\citet{bbb+12}
  reports 21 discoveries with no timing solution; for PSR J1825$-$31
  we do not have enough good timing points to give a coherent timing
  solution yet: the pulsar scintillates and, because of its long
  period, is severely affected by RFI, hence rarely detected. For
  PSR J1816$-$19 (aka J1817$-$1938) a timing solution has been
  presented by \citet{kek+13}, who independently discovered it.} Eight
further pulsars, including 3 MSPs, have been recently discovered in
HTRU-med through a novel pipeline and a neural net candidate selection
algorithm \citep{mbc+19}.

The paper is structured as follows: in section \S \ref{sec:obs} we
describe the observations and data analysis; in \S \ref{sec:time} we
report the results of the timing campaign on 25 HTRU-med pulsars; in
\S \ref{sec:high} we present details of the most interesting sources
of our sample and in \S \ref{sec:cfr} we compare the main
characteristics of all the 113 HTRU-med pulsars with
those of the pulsars found by previous experiments in the same
area. Conclusions are drawn in \S \ref{sec:sum}.

\section{Observations and Data Analysis} 
\label{sec:obs}

The survey observations and the standard search pipeline used for the
discovery of the new pulsars presented in this paper are fully
described in \citet{kjv+10}. Here we briefly summarise the main
points. The 7312 ($\times 13$ beams) 9-minute long HTRU-med pointings
(whose observing date, beam number and galactic coordinates are
reported in the on-line supplementary material) were acquired using
the Parkes 20-cm Multibeam receiver \citep{swb+96} and the BPSR
digital signal processor \citep{kjv+10} over a bandwidth of 400 MHz
(340 MHz, after filtering for known RFI) centered at 1382 (1352)
MHz. The total bandwidth was split into 1024 frequency channels and
the data were 2-bit sampled every 64 $\rm{\mu}$s. This gives $8\times$
better frequency resolution and $2\times$ better time resolution than
the previous highest-time resolution 20-cm Parkes pulsar surveys
\citep{ebvb01,bjd+06,jbo+09,bkl+13}, making HTRU-med particularly sensitive
to rapidly spinning and/or high dispersion measure (DM) pulsars.

After time and frequency domain RFI removal, the data were dedispersed
using 1196 DM values ranging from 0 to 1000 pc cm$^{-3}$. Each
dedispersed time series was then Fourier transformed and the resulting
power spectra were harmonically summed and inspected for significant
peaks. Additionally, a single-pulse search in the time domain (whose
final results will be reported elsewhere) was performed in
parallel. Candidates from all DMs were sorted and the resulting
selection was folded for visual \citep[and/or artificial neural
  network;][]{bbb+12} inspection. Good candidates were finally
reobserved to confirm or reject their astrophysical nature.

Follow-up observations of the new discoveries were initially performed
with the same setup as the survey, using the central beam of the
Multibeam receiver only or, for long-period pulsars, folding on-line
with the ATNF digital filterbank
DFB3\footnote{\tt{http://www.jb.man.ac.uk/pulsar/observing/DFB.pdf}}. A
grid of 5 observations, one at the pulsar position and four 0.15
degrees to the East, West, North and South were first performed in
order to improve the pulsar position. Observations at the improved
position were subsequently obtained at 20 cm either with the Parkes or
Jodrell Bank (Manchester, UK) Lovell telescopes, the latter being used
for pulsars with declination north of $-35^{\circ}$. At both
telescopes the DFB3 backend was mainly used, except for the MSP
follow-up at Jodrell Bank, for which a backend based on ROACH FPGA
boards\footnote{\tt{https://casper.berkeley.edu/wiki/ROACH}}
\citep{bjk+16}, allowing coherent dedispersion, was chosen. A small
number of Parkes observations were performed in parallel also with the
CASPER Parkes Swinburne
Recorder\footnote{\tt{http://astronomy.swin.edu.au/pulsar/?topic=caspsr}}
(CASPSR) and the ATNF Parkes Swinburne
Recorder\footnote{\tt{http://astronomy.swin.edu.au/pulsar/?topic=apsr}}
(APSR) backends in coherent dedispersion mode, for redundancy. All but
the BPSR observations were folded on-line with sub-integration times
ranging from 8 s to 32 s. Details of the different observing
configurations for timing observations are reported in Table
\ref{tab:dfb}.

\begin{table}
\begin{center}
\caption{Main parameters of the observing systems adopted for the
  timing observations. Columns 2 to 5 report the central frequency,
  the bandwidth and the number of frequency channels, while the last
  column indicates whether the data were coherently or incoherenly
  dedispersed.}
\label{tab:dfb}
\begin{tabular}{lcccl}
\hline
Backend & $\nu_c$ (MHz) & BW (MHz) & N$_{\rm{chan}}$ & dedisp \\
\hline
BPSR & 1352 & 340 & 871 & incoh \\ 
DFB3 & 1369 & 256 & 1024 & incoh \\
CASPSR & 1352 & 340 & 871 & coh \\ 
APSR & 1369 & 256 & 512 & coh \\
DFB$_{\rm{JB}}$ & 1532 & 384 & 1024 & incoh \\
ROACH & 1532 & 400 & 1024 & coh \\
\hline
\end{tabular}
\end{center}
\end{table}

For the timing analysis, all data archives were preliminarily cleaned
of RFI using the interactive {\tt{pazi}} software of the
{\tt{psrchive}} package \citep{hvm04}. The archives were then summed
in time and frequency to form a single integrated pulse profile per
observation. The highest signal-to-noise ratio (S/N) profile was
initially adopted as the standard template to obtain the times of
arrival (ToAs) from each observation through cross-correlation using
{\tt{psrchive}}'s {\tt{pat}} application. The ToAs were finally
analysed using {\tt{tempo2}} \citep{hem06} to produce a timing
model. The planetary ephemeris DE421 \citep{fwb09} and the TT(TAI)
time standard (e.g. \citealt{lk05}) were used. All our timing results
are in Barycentric Coordinate Time (TCB). A second iteration of
{\tt{pat}} and {\tt{tempo2}} was run on the archives after they were
updated with the timing model obtained in the previous step, and
using, for the cross-correlation, a new noise-free standard template
created by using {\tt{paas}} to fit one or multiple von Mises components
to a high S/N profile obtained summing all the updated archives.

\section{New discoveries and timing results}
\label{sec:time}

In this section, after briefly presenting the six new pulsars
discovered, we describe the timing results of our follow-up
campaigns. Details on some specific pulsars of interest are discussed
in section \S \ref{sec:high}. Table \ref{tab:new} lists the discovery
parameters of the six new pulsars presented for the first time in this
paper. The S/N of the folded pulse profile in the discovery
observation, the beam in which the pulsar was detected and the offset
of its centre with respect to the final timing position (see Table
\ref{tab:timing}) are reported.

\begin{table}
\begin{center}
\caption{Discovery parameters of the new pulsars. Columns 2 to 4 list,
  for each pulsar, the discovery S/N of the folded profile, the number
  of the beam in which the pulsar was first detected and the
  positional offset between its centre and the timing position of the
  pulsar. The beams are well approximated by a Gaussiam with FWHM of
  $\sim 14'$ \citep{mlc+01}. The discovery parameters for all HTRU-med
  pulsars can be found here
  http://pulsar.oa-cagliari.inaf.it/pulsar/HTRU-med/}
\label{tab:new}
\begin{tabular}{cccc}
\hline
PSR name & S/N & Beam & $\Delta$pos ($'$)\\
\hline
J1518$-$3952 & 38.1 & 01 & 3.3 \\ 
J1629$-$3825 & 16.3 & 05 & 4.7 \\
J1749$-$5417 & 11.1 & 01 & 6.9 \\
J1824$-$0132 & 11.4 & 03 & 4.8 \\
J1826$-$2415 & 10.3 & 04 & 5.9 \\
J1835$-$0847 & 18.6 & 09 & 1.8 \\
\hline
\end{tabular}
\end{center}
\end{table}

For Galactic latitudes between $\pm5^{\circ}$ and $\pm15^{\circ}$,
where all new discoveries except PSR J1835$-$0847 are located, the
previous deepest 20-cm search experiment was the Swinburne
intermediate-latitude pulsar survey \citep[SWIN$_{\rm{MED}}$; ][]{ebvb01}, with
half the integration time per pointing and 8$\times$ worse frequency
resolution than HTRU-med. In most cases, the reason why the pulsars in
Table \ref{tab:new} were not previously detected, likely lies in their
intrinsically low flux densities: usually search candidates with S/N
smaller than $\sim10$ are not visually inspected, or are hard to
classify as credible pulsars, hence could have easily been missed. For
MSP J1826$-$2415, moreover, the DM smearing expected from the
Swinburne survey is 0.8 ms, a sufficiently large value, compared to
the pulsar's spin period, to further reduce the pulse S/N. PSR
J1518$-$3952, on the other hand, has a high discovery S/N and is
clearly visible in all our 3-minute long timing observations. Folding
the closest SWIN$_{\rm{MED}}$  pointing (whose position is coincident with our
discovery pointing) with our timing parameters (see
Tab. \ref{tab:timing}), indeed, results in a detection with a S/N of
17.5. The vicinity of the pulsar period to 0.5 s (a
sub-harmonic of the 50 Hz power line signal) might be the reason why
the signal was not considered as a credible candidate.

The only pulsar whose position is not covered by SWIN$_{\rm{MED}}$ is
J1835$-$0847. The closest observation to its position covered by
Parkes surveys is a 35-minute long 20-cm observation taken as part of
the Parkes Multibeam Pulsar Survey \citep[PMPS; ][]{mlc+01} 7.8$'$
away from the pulsar position. Considering the coarser time
(4$\times$) and frequency (8$\times$) resolution, the 1-bit sampling
and, most importantly, the positional offset, PSR J1835$-$0847 should
have, in PMPS data, a S/N of $\sim10$, consistent with its
non-detection. The positions of PSR J1824$-$0132 and MSP J1826$-$2415
are also covered by the outer PMPS pointings.  Despite the longer
integration time, the positional offsets with respect to the centre of
the closest beam and the 3-MHz channels, make the expected S/N too low
for a firm detection.

Folding the archival data\footnote{obtained through the Parkes Data
  Access
  Portal\newline\indent{\tt{https://data.csiro.au/dap/public/atnf/pulsarSearch.zul}}}
closest to our new pulsar positions with the ephemeris we obtained in
our timing campaign (see Table \ref{tab:timing}), resulted indeed,
besides the aforementioned case of J1518$-$3952, in only faint
detections of J1835$-$0847 (S/N = 10.1 after RFI removal),
J1826$-$2415 (S/N = 9.5) and J1824$-$0132 (S/N = 10.3) in PMPS data,
and of J1629$-$3825 in SWIN$_{\rm{MED}}$ (S/N = 10.9) and the Parkes 70-cm all-sky
survey \citep[S/N = 12.2; ][]{mld+96}. These low S/N detections from
archival data are not included in our timing solutions, due to unknown
instrumental offsets.

Timing observation of the pulsars presented here spanned, at Parkes,
on average 400 days for each pulsar, except for the nulling PSR
J1638$-$4233 (\S \ref{sec:1638}), observed for twice as long because
of the paucity of detections. Lovell telescope observations, on the
other hand, continued for longer, and in most cases are still ongoing
as of August 2018.

In Table \ref{tab:timing} we list the main timing parameters for the
latest 25 pulsars (6 new and 19 whose discovery was reported in
\citealt{bbb+12}) discovered in HTRU-med through our standard
FFT-based, non-accelerated pipeline, while in Table \ref{tab:derived}
we list the parameters derived from the timing analysis, along with
the widths of the pulse profiles and the pulsars mean flux densities
at 20-cm. For the pulsar timed at Jodrell Bank (with declinations
above $-35^{\circ}$), the flux densities are estimated averaging the
values obtained through the radiometer equation
(e.g. \citealt{mlc+01}) for each single observation, and the errors
are their standard deviation. Given the fact that timing observations
with the Lovell telescope were done at low elevation, gain loss and
spill-over effects have also been taken into account. Parkes' values,
intsead, are obtained through calibrated data using psrchive's
{\tt{pac}}.

\renewcommand{\tabcolsep}{3pt}
\begin{table*}
\begin{center}
\caption{Main timing parameters for 25 HTRU-med pulsars. Columns 1 and
  2 report the pulsar name and the provisional name used in
  \citet{bbb+12}. Columns 3 to 8 report, J2000 right ascention and
  declination, spin period and spin period first derivative, the epoch
  at which the period is referred and the dispersion measure. The
  three last columns report the data span of the timing observations,
  the rms of the timing residuals, and the number of ToAs. Numbers in
  parentheses are the 2$\sigma$ errors on the last quoted
  digit(s).\newline$^\dag$For this pulsar we report the pre-glitch
  parameters and rms, but the total number of ToAs acquired over the
  entire data-span. \newline$^\ddag$For these pulsars also a spin
  frequency second derivative was fitted (see \S\ref{sec:noise}).}
\label{tab:timing}
\begin{tabular}{@{}lcllllcccrr}
\hline
 \multicolumn{1}{c}{PSR name} & Alias & \multicolumn{1}{c}{RAJ} & \multicolumn{1}{c}{DECJ} & \multicolumn{1}{c}{P} & \multicolumn{1}{c}{$\dot{\rm{P}}$} & EPOCH & DM & Data span & \multicolumn{1}{c}{rms} & \multicolumn{1}{c}{N(ToA)} \\ 
     &  & \multicolumn{1}{c}{(h:m:s)} & \multicolumn{1}{c}{(d:m:s)} & \multicolumn{1}{c}{(s)} & \multicolumn{1}{c}{($10^{-15}$)} & (MJD) & (pc/cm$^3$) & (MJD) & \multicolumn{1}{c}{(ms)} & \\
\hline
J0836$-$4233 & J0835$-$42 & 08:36:05.85(3)   & $-$42:33:24.9(3)  & 0.73843560627(4) & 1.320(4) & 56290 & 180 & 55954--56626 & 0.64 & 26 \\
J1105$-$4353 & J1105$-$43 & 11:05:24.84(11)  & $-$43:53:05(3)    & 0.35111186198(7) & 2.50(4) & 55976 & 45 & 55804--56147 & 0.94 & 14 \\
J1132$-$4700 & J1132$-$46 & 11:32:14.52(3)   & $-$47:00:36.5(3)  & 0.325633290107(6) & 0.7768(7) & 56122 & 123 & 55756--56488 & 0.63 & 36 \\
J1518$-$3952 & --         & 15:18:56.968(18) & $-$39:52:30.3(7)  & 0.499195890475(15) & 0.237(8) & 56442 & 101 & 56256--56627 & 0.27 & 25 \\
J1530$-$6343 & J1530$-$63 & 15:30:47.608(20) & $-$63:43:13.6(3)  & 0.910319586358(11) & 0.825(7) & 56093 & 206 & 55906--56279 & 0.19 & 28 \\
J1551$-$6214 & J1552$-$62 & 15:51:36.788(8)  & $-$62:14:23.37(9) & 0.1988386042970(8) & 0.0228(5) & 56108 & 122 & 55905--56310 & 0.08 & 24 \\
J1614$-$3846 & J1614$-$38 & 16:14:42.52(8)   & $-$38:46:35(6)    & 0.46410631422(9) & 0.92(4) & 55876 & 111 & 55698--56054 & 1.33 & 17 \\
J1629$-$3825 & --         & 16:29:14.618(11) & $-$38:25:29.3(5)  & 0.526364586499(9) & 0.4438(14) & 56376 & 127 & 56125--56626 & 0.37 & 34 \\
J1636$-$2614$^\dag$ & J1635$-$26 & 16:36:11.506(17) & $-$26:14:47.0(13) & 0.510453755610(8) & 4.0580(11) & 57018 & 94 & 55910--58127 & 0.29 & 95 \\
J1638$-$4233 & J1638$-$42 & 16:38:24.71(6)   & $-$42:33:55(3)    & 0.51092936864(3) & 10.088(3) & 56253 & 406 & 55834--56672 & 0.86 & 9 \\
J1704$-$5236 & J1705$-$52 & 17:04:40.569(15) & $-$52:36:57.4(5)  & 0.230708491737(3) & 0.05.2(3) & 56108 & 170 & 55905--56311 & 0.30 & 24 \\
J1719$-$2330$^\ddag$  & J1719$-$23 & 17:19:36.1(4)    & $-$23:30:14(72)   & 0.453992694395(19) & 2.1861(4) & 56614 & 101 & 55690--57848 & 3.15 & 75 \\
J1749$-$5417 & --         & 17:49:47.06(3)   & $-$54:17:30.7(3)  & 0.307576751860(6) & 0.1496(7) & 56388 & 69 & 56066--56709 & 0.44 & 23 \\
J1757$-$1500$^\ddag$  & J1757$-$15 & 17:57:12.752(17) & $-$15:00:28.5(17) & 0.179354019096(13) & 1.5039(4) & 57038 & 150 & 55955--58121 & 1.82 & 128 \\
J1802$-$0523 & J1802$-$05 & 18:02:12.14(8)   & $-$05:23:29(4)    & 1.68057296916(9) & 2.649(3) & 56908 & 121 & 55696--58119 & 8.95 & 104 \\
J1818$-$0151 & J1818$-$01 & 18:18:25.77(3)   & $-$01:51:54.0(11) & 0.837534946983(11) & 9.1061(5) & 57015 & 209 & 55909--58120 & 2.39 & 78 \\
J1824$-$0132 & --         & 18:24:55.575(9)  & $-$01:32:24.0(4)  & 0.223728519703(4) & 0.02.82(3) & 57658 & 79 & 57182--58133 & 0.67 & 43 \\
J1826$-$2415 & --         & 18:26:36.0260(9) & $-$24:15:54.5(3)  & 0.0046957571305605(7) & 0.00001736(3) & 57054 & 81 & 55957--58152 & 0.03 & 97 \\
J1835$-$0847 & --         & 18:35:45.72(5)   & $-$08:47:20(3)    & 0.84649413462(6) & 0.485(4) & 57653 & 850 & 57183--58123 & 2.83 & 45 \\
J1837$-$0822 & J1837$-$08 & 18:37:39.654(10) & $-$08:22:12.4(7)  & 1.099195959243(8) & 0.1209(3) & 57036 & 506 & 55955--58117 & 1.10 & 122 \\
J1840$-$0445$^\ddag$  & J1840$-$04 & 18:40:45.5(3)    & $-$04:45:08(13)   & 0.42231632097(13) & 11.280(3) & 57053 & 379 & 55955--58151 & 28.32 & 94 \\
J1900$-$0933 & J1900$-$09 & 19:00:31.19(3)   & $-$09:33:48.1(15) & 1.423889144409(19) & 0.2263(9) & 56979 & 189 & 55840--58118 & 2.43 & 103 \\
J1902$-$1036$^\ddag$  & J1902$-$10 & 19:02:42.399(10) & $-$10:36:13.0(7)  & 0.786813538431(8) & 7.53681(19) & 56916 & 91 & 55699--58124 & 0.99 & 95 \\
J1904$-$1629 & J1904$-$16 & 19:04:51.37(3)   & $-$16:29:19(3)    & 1.541412044233(20) & 3.1900(8) & 57039 & 145 & 55956--58119 & 1.45 & 58 \\
J1920$-$0950 & J1920$-$09 & 19:20:55.93(3)   & $-$09:50:01.0(18) & 1.037824001161(19) & 0.3928(7) & 57037 & 93 & 55955--58119 & 2.66 & 79 \\
\hline
\end{tabular}
\end{center}
\end{table*}

\renewcommand{\tabcolsep}{6pt}
\begin{table*}
\begin{center}
\caption{Derived parameters and other measured parameters for 25
  HTRU-med pulsars. Column 2 and 3 report the Galactic longitude and
  latitude; columns 4 and 5 the DM-derived distances obtained using
  the NE2001 \citep{cl02} and YMW16 \citep{ymw17} models for the
  distribution of free electrons in the Galaxy, respectively. We note
  that the values obtained by the two models are, in some cases,
  significantly different, with the YMW17 model often giving very
  large distances. Columns 6 to 9 list the dipolar surface magnetic
  field, the spin down energy and the characteristic age of the
  pulsars. The last 3 columns report the pulse width at 50\% and 10\%
  of the pulse height, and the pulsar mean flux density at 1.4
  GHz.\newline$^{\dag}$ The values
  reported in the table are for the main peak only. The 50\% and 10\%
  widths of the secondary peak are 0.5(2) ms and 1.8(8) ms
  respectively.}
\label{tab:derived}
\begin{tabular}{lrrcccccccc}
\hline
\multicolumn{1}{c}{PSR name} & \multicolumn{1}{c}{$l$} & \multicolumn{1}{c}{$b$} & D$_{\rm{DM}}^{\rm{ne2001}}$ &
 D$_{\rm{DM}}^{\rm{ymw16}}$ & B$_{\rm{surf}}$ & $\dot{\rm{E}}$ &
 $\tau_c$ & W50  & W10 & S$_{1400}$\\
 & \multicolumn{1}{c}{($^{\circ}$)} & \multicolumn{1}{c}{($^{\circ}$)}
 & (kpc) & (kpc) & ($10^{11}$ G) &
 ($10^{32}$ erg/s) & ($10^6$ yr) & (ms) & (ms) & (mJy)\\
\hline
J0836$-$4233  &  261.537  &  $-$1.107  &  0.5  &  3.6  &  9.9  &   1.3 & 8.9 &  14.21(1) &  25.91(1) & 0.129(3) \\ 
J1105$-$4353  &  283.486  &  14.947    &  1.6  &  0.1  &  9.5  &  22.8 & 2.2 & 13.107(5) & 23.923(5) & 0.189(6) \\ 
J1132$-$4700  &  289.177  &  13.736    &  5.2  &  2.7  &  5.1  &   8.9 & 6.6 &   17.6(5) &   29.0(3) & 0.490(5) \\ 
J1518$-$3952  &  331.368  &  14.715    &  3.2  &  11.4$\;\,$  &  3.5  &  0.8 &  33.4 & 7.6(1) & 55.2(6) & 0.90(1) \\ 
J1530$-$6343  &  319.506  &  $-$6.101  &  5.0  &  11.5$\;\,$  &  8.8  &  0.4  &  17.5 & 11(5) & 32.6(3) & 0.37(9) \\ 
J1551$-$6214  &  322.300  &  $-$6.350  &  2.7  &  3.9  &  0.7  &  1.1  &  138.0 & 2.8(1)  & 6.46(6) & 0.312(5) \\ 
J1614$-$3846  &  340.628  &  8.788     &  2.7  &  5.7  &  6.6  &  3.7  &  7.9 & 24.471(6) & 44.726(6) & 0.182(6)\\ 
J1629$-$3825  &  342.889  &  7.006     &  2.9  &  5.6  &  4.9  &  1.2  &  18.8 & 7.0(3) & 14.8(3) & 0.383(4)\\ 
J1636$-$2614  &  353.120  &  14.036    &  2.9  &  8.3  &  14.6  &  12.0  &  1.9 &  7.273(4) & 13.271(7) & 0.19(5)\\ 
J1638$-$4233  &  341.007  &  2.924     &  7.6  &  22.2$\;\,$  &  23.0  &  29.9  &  0.8 & 18(1) & 44(3) & 0.35(1)\\ 
J1704$-$5236  &  336.051  &  $-$6.869  &  4.0  &  9.6  &  1.1  &  1.7  &  70.1 & 15.3(3) & 24.1(5) & 0.605(9) \\ 
J1719$-$2330  &  1.402    &  7.846     &  2.4  &  4.3  &  10.1  &  9.2 &  3.3 & 8.9(6) & 18.0(4) & 0.16(6)\\ 
J1749$-$5417  &  338.275  &  $-$13.427  &  1.8  &  3.3  &  2.2  &  2.0  &  32.6 & 12.133(3) & 22.153(4) & 0.321(9)\\ 
J1757$-$1500  &  13.367  &  4.785      &  3.3  &  4.3  &  5.3  &  103.0  &  1.9 & 2.1(7) & 4.78(5) & 0.09(2)\\ 
J1802$-$0523  &  22.425  &  8.398      &  3.4  &  4.1  &  21.4  &  0.2  &  10.1 & 64.4(5) & 146(3) & 0.15(7)\\ 
J1818$-$0151  &  27.482  &  6.479      &  6.2  &  19.3$\;\,$  &  27.9  &  6.1  &  1.5 & 15(1) & 35.1(9) & 0.09(2)\\ 
J1824$-$0132  &  28.525  &  5.188      &  2.3  &  3.1  &  0.8  &  0.9  &  126.0 & 7.4(2) & 22.4(6) & 0.12(5)\\ 
J1826$-$2415  &  8.553  &  $-$5.718    &  2.3  &  2.8  &  0.003  &  66.2  &  4290 & 0.361(1)$^{\dag}$ & 1.9(9)$^{\dag}$ & 0.28(9)\\ 
J1835$-$0847  &  23.331  &  $-$0.548   &  9.4  &  6.2  &  6.5  &  0.3  &  27.6 & 46(1) & 108(5) & 0.4(1)\\ 
J1837$-$0822  &  23.918  &  $-$0.772   &  6.7  &  5.2  &  3.7  &  0.03  &  144.0 & 37(3) & 53.7(3) & 0.2(2)\\ 
J1840$-$0445  &  27.486  &  0.203      &  6.0  &  4.6  &  22.1  &  59.1  &  0.6 & 22.4(7) & 53(2) & 0.37(8)\\ 
J1900$-$0933  &  25.413  &  $-$6.346   &  5.2  &  13.7$\;\,$  &  5.7  &  0.03  &  99.7 & 51(19) & 250(2) & 0.4(2)\\ 
J1902$-$1036  &  24.716  &  $-$7.291   &  2.6  &  3.8  &  24.6  &  6.1  &  1.7 & 11.80(1) & 21.528(7) & 0.13(7) \\ 
J1904$-$1629  &  19.590  &  $-$10.332  &  4.6  &  25.0$\;\,$  &  22.4  &  0.3  &  7.7 & 12.4(9) & 30.8(4) & 0.17(7)\\ 
J1920$-$0950  &  27.417  &  $-$10.977  &  3.0  &  6.2  &  6.5  &  0.1 &  41.9 & 39.2(6) & 60(1) & 0.20(8) \\ 
\hline
\end{tabular}
\end{center}
\end{table*}

\begin{figure}
\includegraphics[width=8.5cm, angle=270]{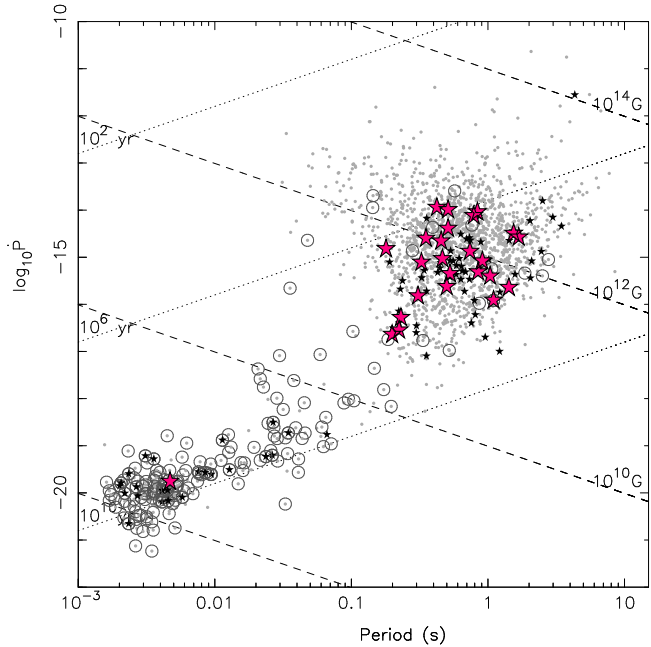}
\caption{Period-period derivative diagram. Large pink stars are the 25
  pulsars presented in this work. Small black stars are the other
  HTRU-med pulsars discovered with our standard pipeline. Grey dots
  are previously known Galactic field radio pulsars. Binary pulsars are
  surrounded by a grey circle. Dashed and dotted lines are equal
  magnetic field and equal characteristic age lines,
  respectively. Data were obtained from the ATNF Pulsar Catalogue
  \citep{mhth05} v. 1.58.}
\label{fig:ppdot}
\end{figure}

The 25 pulsars in our sample are shown as large stars in the
period-period derivative diagram of Figure \ref{fig:ppdot}, together
with the other previously published HTRU-med pulsars (small black stars), and
the rest of the known Galactic field radio pulsar population (grey
dots) as derived from version 1.58 of the ATNF Pulsar
Catalogue\footnote{\tt{http://www.atnf.csiro.au/people/pulsar/psrcat/}}
\citep{mhth05}

\begin{figure*}
\includegraphics[width=18cm]{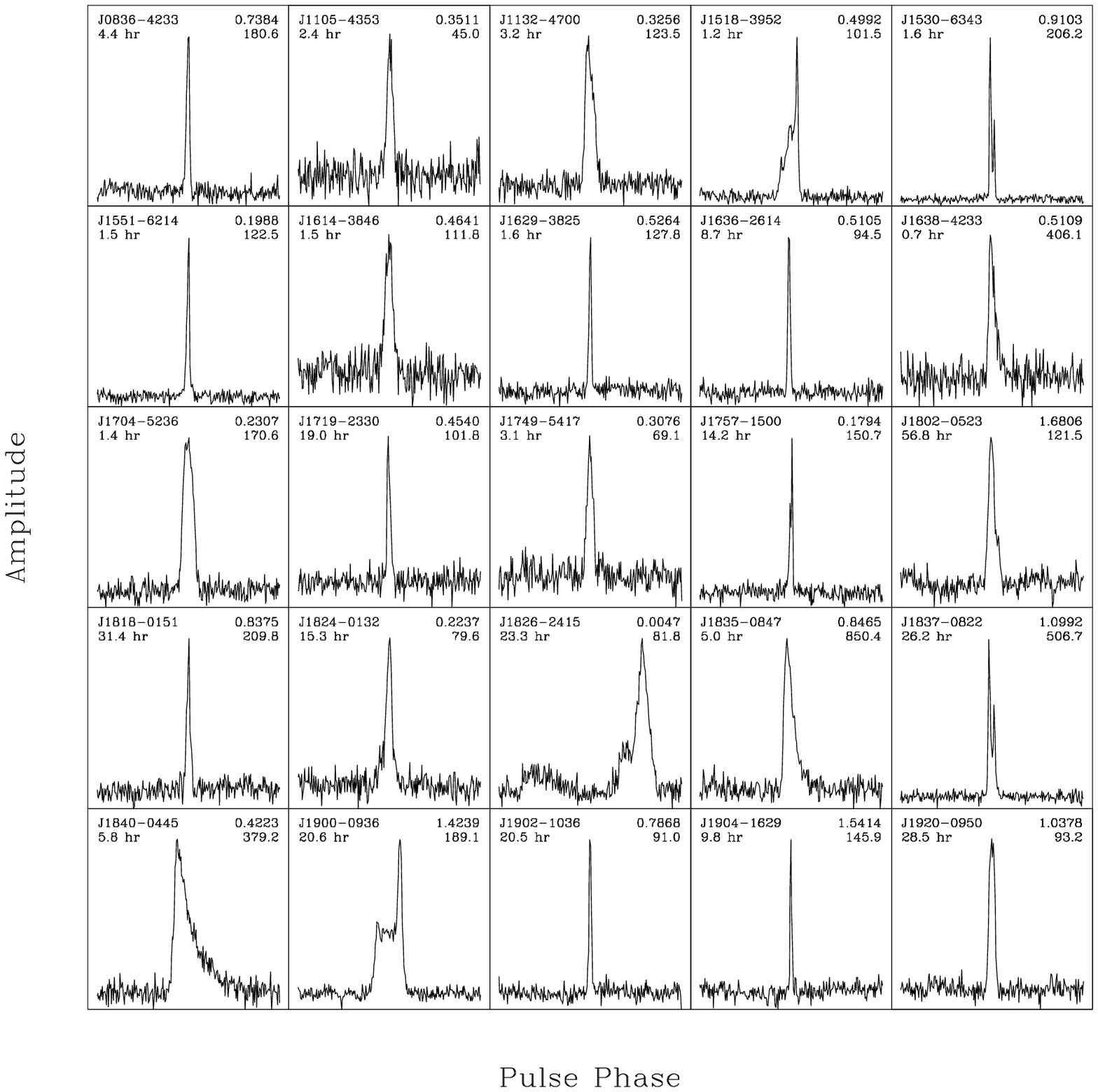}
\caption{Integrated pulse profiles for the 25 pulsars presented
  here. For each pulsar, in the top left corner we report the source
  name and the integration time contributing to the pulse profile, and
  in the top right corner the spin period in seconds and the disperion
  measure in pc/cm$^3$. The entire pulse period is shown.}
\label{fig:prof}
\end{figure*}

Figure \ref{fig:prof} shows the integrated pulse profiles for our
sample of pulsars, obtained by summing, for each source, the entire 20
cm dataset used for timing (see Table \ref{tab:timing}). The total
summed integration time contributing to each profile is shown in the
top-left corner of the plot. Observations are uncalibrated, hence the
y-axis is in arbitrary units. For the pulsars observed at Parkes, the
data were folded over 512 time bins, while for those observed at
Jodrell Bank 1024 bins were used. In both cases the profile bins were
reduced to 256 for plotting purposes only. DM smearing is smaller than
the shown time bin in all cases.

\section{Highlights}
\label{sec:high}

In the following sub-sections we discuss in more detail the most
interesting features of some of the pulsars presented in this paper.

\subsection{PSR J1837$-$0822: a long-period pulsar in a 98-day orbit}
\label{sec:longP}

PSR J1837$-$0822 is a 1.1-s pulsar in a 98-day, mildly eccentric orbit
(e = 0.024; see Table \ref{tab:J1837}) around a low-mass companion
star. This pulsar is one of only 21 known long-period (P $>$ 0.1 s)
pulsars in a binary system in the Galactic field.

The minimum mass of the companion, derived from the mass function
assuming a 1.35 M$_{\odot}$ neutron star (NS), is 0.27 M$_{\odot}$ for
an edge-on orbit. The median companion mass, calculated for an
inclination angle of 60$^{\circ}$, would be 0.32 M$_{\odot}$, and
could be as high as 0.41 M$_{\odot}$ for a pulsar mass of 2
M$_{\odot}$ \citep[the highest NS mass precisely
measured;][]{afw+13}. A 90\% confidence upper limit on the companion
mass can be obtained using an orbital inclination angle of
26$^{\circ}$: depending on the NS mass (1.35 or 2 M$_{\odot}$) the
maximum mass for PSR J1837$-$0822's companion ranges from 0.73 to 0.9
M$_{\odot}$.

\begin{table}
\begin{center}
\caption{Orbital parameters for the binary long-period pulsar
  J1837$-$0822. The BT binary model \citep{bt76} was adopted. Row 1 to
  6 list the orbital period, the projected semi-major axis, the epoch
  and the longitude of the periastron, the eccentricity and the mass
  function. The minimum (M2$_{\rm{min}}$), median (M2$_{\rm{med}}$) and
  maximum (M2$_{\rm{max}}$) companion masses (see text) were obtained
  assuming a 1.35 M$_{\odot}$ NS mass. 2$\sigma$ errors on the last
  quoted digit(s) are reported in parentheses.}
\label{tab:J1837}
\begin{tabular}{ll}
\hline
\multicolumn{2}{c}{PSR J1837$-$0822}\\
\hline
$P_{\rm{b}}$ (d) & 98.36371(3) \\ 
$a1$ (lt-s) & 40.8780(4) \\
T0 (MJD) & 55978.1170(10) \\
$\omega$ (deg) & 21.94(4) \\
e & 0.024319(15) \\
$f_{\rm{m}}$ (M$_{\odot}$) & 0.0075802(2) \\
M2$_{\rm{min}}$ (M$_{\odot}$) & 0.27 \\
M2$_{\rm{med}}$ (M$_{\odot}$) & 0.32 \\
M2$_{\rm{max}}$ (M$_{\odot}$) & 0.73 \\
\hline 
\end{tabular}
\end{center}
\end{table}

\begin{figure}
\includegraphics[width=8.5cm]{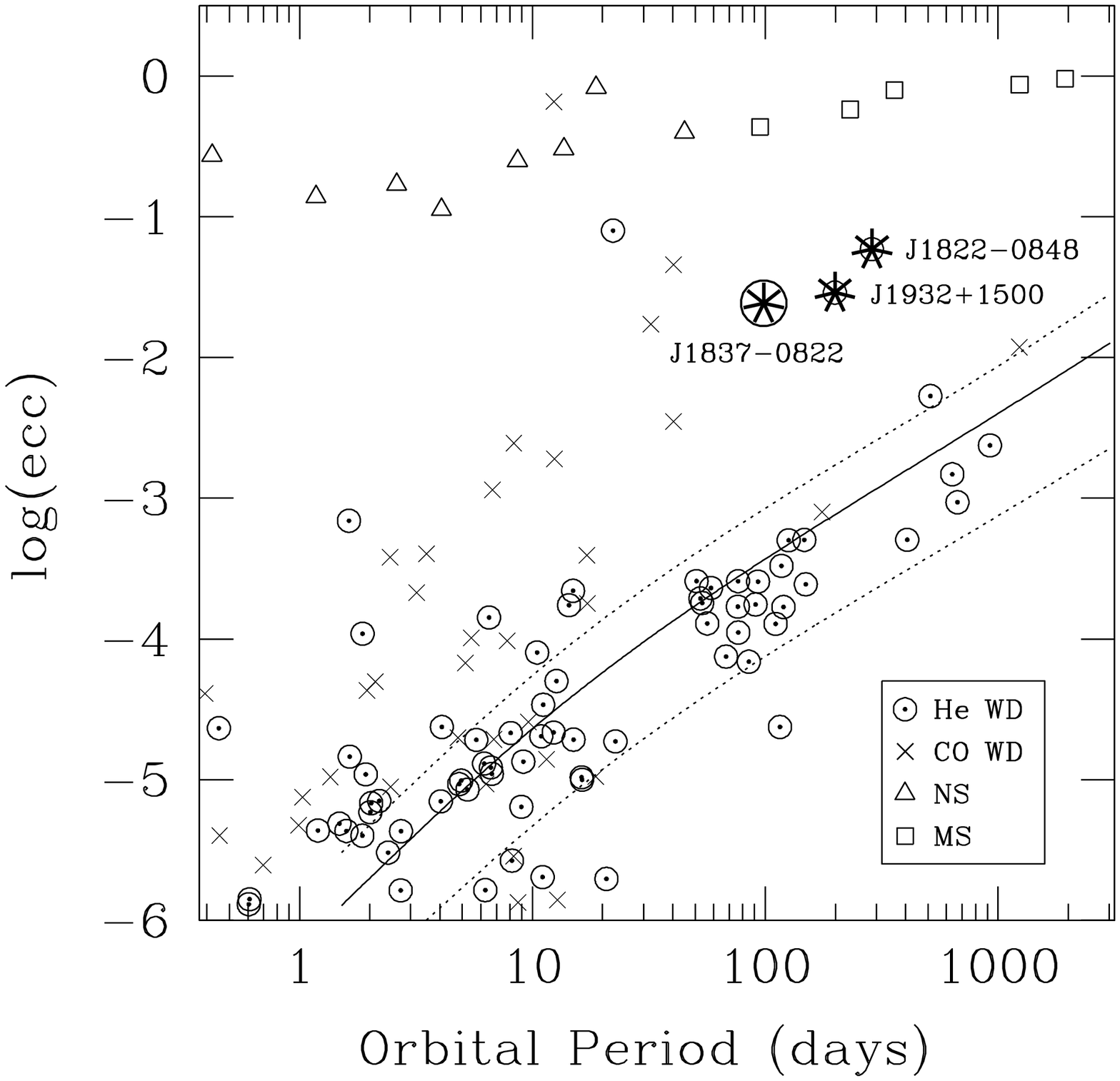}
\caption{Orbital eccentricity versus orbital period of binary pulsars
  in the Galactic field. Circled dots are pulsars with He-WD
  companions, crosses have Carbon-Oxigen companions (CO-WD), while
  triangles and squares denote pulsars with NS and main sequence (MS)
  companions respectively (from the ATNF pulsar catalogue;
  \citealt{mhth05}). According to the model of \citet{pk94}, 95\% of
  recycled pulsars with low mass He-WD companions should fall within
  the dotted lines. The large asterisks are the three long period
  binary pulsars J1837$-$0822, J1932$+$1500 and J1822$-$0848.}
\label{fig:pbecc}
\end{figure}

The intermediate eccentricity of this system, together with its long
spin and orbital periods and its relatively small companion mass (see
Tab \ref{tab:J1837}), likely makes it the third example of a weakly
(if at all) recycled long-period pulsar \citep{tlk12}, together with
J1932$+$1500 \citep{lsb+17} and J1822$-$0848 \citep{lfl+06}. Similarly
to the other two, the eccentricity of PSR J1837$-$0822's orbit is too
small for the pulsar to be a non-recycled object relatively recently
born in a binary system like e.g. PSR J2032$+$4127 \citep{lsk+15} or
PSR J1740$-$3052 \citep{lfl+06}, in eccentric orbits around massive
main sequence stars, or like PSR J1141$-$6545 \citep{ts00}, a NS born
after its WD companion. On the other hand, the eccentricity is too
high for the system to be similar to mildly-recycled pulsars, having
typically lower spin periods ($\lsim 0.1$ s) and magnetic fields
($\lsim 10^{10}$ G). As further evidence of its peculiarity, PSR
J1837$-$0822 does not follow the orbital period-eccentricity relation
of \citet{phi92} for Helium WD (He-WD) companions and falls close to
the other two members of this new small family, in the $P_{\rm{b}}$-e
plot (see Fig. \ref{fig:pbecc}).

The position of these three pulsars in the plot, together with their
long spin periods (distinguishing them, e.g., from the close-by CO-WD
systems, with almost 2 orders of magnitude smaller rotational periods),
puts them in a peculiar category of binary pulsar whose companions
have likely evolved into He-WDs via mass transfer, but in which
accretion onto the NS has not taken place in an effective way, or at
all \citep{tlk12}.

\subsection{PSR J1826$-$2415 and the origin of isolated MSPs}
\label{sec:MSP}

Isolated fully recycled MSPs (defined here as pulsars with spin
periods smaller than 10 ms and magnetic fields smaller than $10^{9}$
G) are relatively rare in the Galactic field. The ATNF pulsar
catalogue lists only 32 such objects. Another 23 MSPs with no orbital
period measured are listed in the catalogue; these however, have no
fully coherent timing solutions yet (only the discovery parameters
with a small number of significant digits are listed), and/or they
do not have a measured spin period derivative yet. Because of this, we
cannot be certain that they do not have a companion star, at the
moment, and in the following we define them as isolated candidates.
PSR J1826$-$2415, a 4.7-ms isolated pulsar, adds to this small and
interesting family of objects whose origin is still under debate. 

In the standard scenario, isolated MSPs are the end product of
binary systems where the pulsars have completely ablated their
companion stars (e.g. \citealt{acrs82,vv88, bv91}). In this framework,
the so-called Black Widows and Redbacks (hereafter ``spiders''),
binary systems where the radio signal from the MSP is eclipsed for a
significant fraction of the orbit by matter stripped from their
companions (see e.g. \citealt{rob13}), would be the progenitors of
isolated MSPs. Whether or not the timescale for complete ablation of
the companion star is short enough, however, is still debated (e.g.
\citealt{kn88,rst89a,sbl+96,ccth13}).

Because of pulsar spin-down, if this model is correct, one would
expect isolated MSP to have longer spin periods $P$ than the spiders
from which they derive. $P$ distributions of Redbacks and Black Widows
(both counted separately and all togehter) have indeed mean and median
values $\sim 2$ ms smaller than those of isolated MSPs (see Table
\ref{tab:Pmean}), supporting the standard scenario. The numbers in
\ref{tab:Pmean} are for Galactic field objects only, since the
evolutionary history of the systems in Globular Clusters could be
dramatically different \citep[e.g][]{vvvr87}.
\begin{table}
\begin{center}
\caption{Mean and median spin periods (columns 3 and 4) for isolated
  MSPs (bona fide ones, first line, and including candidates, second
  line), Redbacks, Black Widows, and spiders all together. Column 2
  lists the number of pulsars in each category.}
\label{tab:Pmean}
\begin{tabular}{lcccc}
\hline
Type & \#PSRs & P$_{\rm{mean}}$ (ms) & P$_{\rm{med}}$ (ms) \\
\hline
Isolated & 32 & 4.47 & 4.72  \\ 
Isolated + cands & 55 & 4.59 & 4.39 \\
Redbacks & 12 & 3.15 & 2.75 \\
Black Widows & 22 &  2.58 & 2.32 \\
Spiders & 34 & 2.79 & 2.48 \\
\hline
\end{tabular}
\end{center}
\end{table}

An observational parameter that, on the other hand, we would expect to
be similar in isolated MSPs and spiders, if in fact they had a commom
origin, is their distance from the Galactic plane Z, a parameter
linked to the natal kick velocity of pulsars, hence to their formation
path. Looking at the average values of Z for the various classes of
Galactic field MSPs, instead, we see that spiders are, on average,
further from the plane (with a mean vertical distance
Z$_{\rm{mean}}=0.8$ kpc), while isolated MSPs appear on average lower
(Z$_{\rm{mean}}=0.3$ kpc) and more similarly distributed to non
eclipsing binaries (Z$_{\rm{mean}}=0.4$ kpc{\footnote{The numbers
above are calculated in PSRcat using the distance from the
\citet{ymw17} electron density model. Compatible results, with spiders
having Z$_{\rm{mean}}$ twice as big as that for isolated and non
eclipsing binaries, are obtained using the \citet{cl02}
model.}}). This seems to argue against a common origin between spider
MSPs and the bulk of the isolated ones (although we must note that,
deriving from the quite uncertain DM-inferred distances, Z values may
have large errors).  Alternative scenarios, such as accretion-induced
collaspe of a white dwarf \citep{bg90}, or white dwarf mergers
\citep{mic82}, appear to be more in line with a low velocity at birth
(hence possibly a small distance from the Galactic plane) for isolated
MSPs; a high velocity and a high altitude are expected for tightly
bound systems \citep{tb96} such as Redbacks and Black Widows, and such
as isolated MSPs would need to have been, if they had to be able to
efficiently ablate their companions.

\subsection{The nulling pulsar J1638$-$4233}
\label{sec:1638}

\begin{figure}
\includegraphics[width=8.5cm]{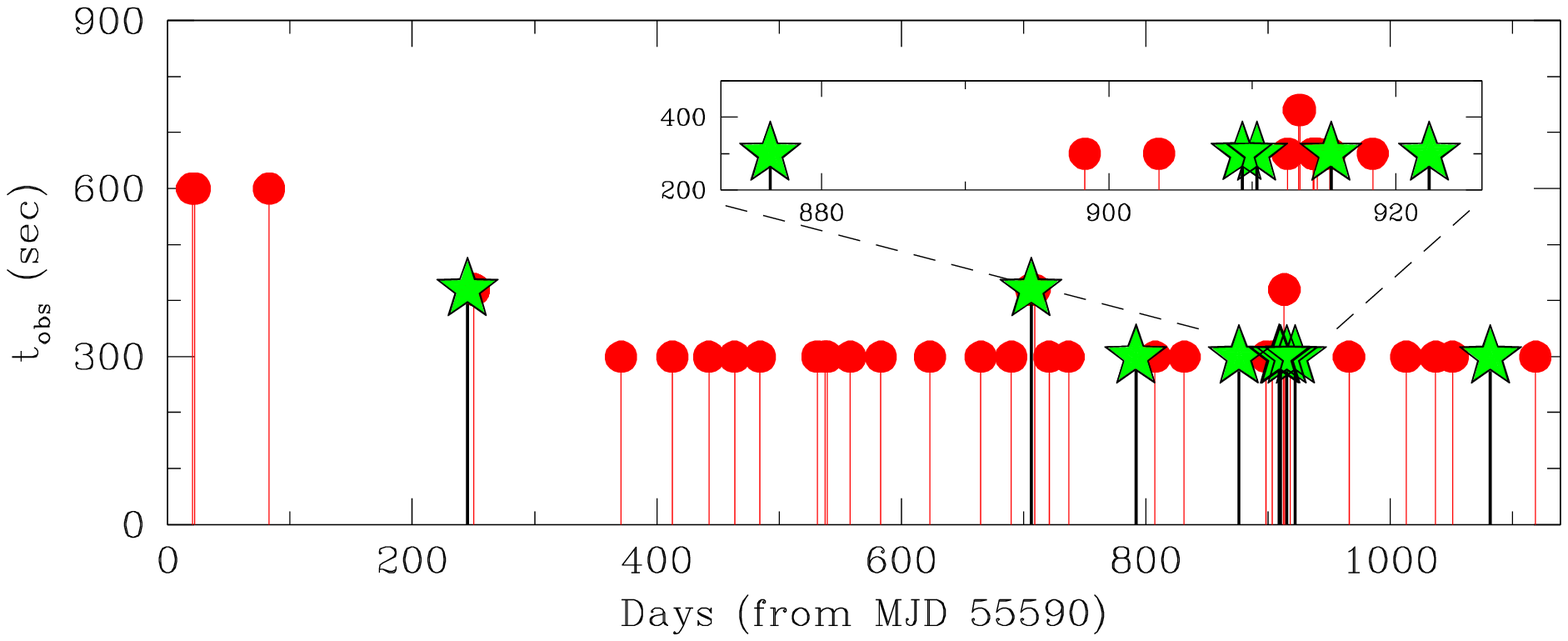}
\caption{Detections (green stars) and non detections (red circles) of
  PSR J1638$-$4233 plotted in a time (days from MJD 55590) vs
  observation length plot. The pulsar was seen on only 9 occasions. The
  inset is a blow-up the MJD range 56463-56516, when more than half of
  the detections were made.}
\label{fig:null}
\end{figure}

PSR J1638$-$4233 was observed at Parkes 52 times, for a total of 4.9
hours over a period of 3 years. Single observations typically lasted 5
minutes. In total, J1638$-$4233 was seen (S/N $> 7$) on only 9
occasions. Figure \ref{fig:null} shows the number and time of
detections for this pulsar.

Even when the pulsar is visible, its pulses appear to switch off (or
null; \citealt{bac70}) for times extending from a single sub-integration
(20 or 30 second, depending on the time resolution of each
observation) to a few minutes: the longest gap between two ON phases
in a single observation is 2 minutes, while the longest observation in
which the pulsar is always off is 10 minutes. To better estimate the
duration of the nulling phases of the observations in which the pulsar
is seen, we cross-correlated the pulsar's standard profile with all
the sub-integrations in the data files, and counted as OFF all the
sub-integrations for which the time of arrival was more than 0.1 pulsar
rotations off with respect to zero phase predicted on the basis of
the timing solution presented in \S \ref{sec:time}. About one third of the
sub-integrations appear to be off. Considering this and the
non-detections, the total nulling fraction is about 90\%, among the
highest known (see e.g. \citealt{wmj07} and \citealt{gaj17}), with the
pulsar ON for only about 33 minutes of the 4.9 observed hours.

Given that more than half of the detections occurred in a short
time-span (5 over 45 days, 4 of which over 13 days), it could also be
that PSR J1638$-$4233 is both a nulling and an intermittent pulsar
\citep{klo+06,crc+12,llm+12,sjmg13}, switching off for long periods,
and being a nuller with a null fraction of about 50\%, in its active
state. In its longer-term behaviour it could be similar to PSRs
J1910$+$0517 and J1929$+$1357 \citep{lsf+17} which show extreme
bimodal long-term intermittency. The latter shows changes in the
spin-down rate which is correlated with a change in the fraction of
time it was on \citep{lsf+17}. Unfortunately the detections are too
sparse to check if there are correlated spin-down changes in PSR
J1638$-$4233, but increased monitoring, perhaps by MeerKAT \citep{jon09}, the
recently opened South African precursor of the Square Kilometre Array,
could reveal if it is indeed a member of the intermittent class.

\subsection{Mode changes in PSR J1757$-$1500}
\label{sec:mode}

Sometimes linked to nulling, mode changing (e.g. \citet{wmj07,lhk+10})
is a phenomenon, observed now in a few tens of pulsars, where the
integrated pulse profile switches between two or more quasi-stable
states (in the case of nulling, in the second state the pulse is off).

\begin{figure}
\includegraphics[width=8.5cm,clip]{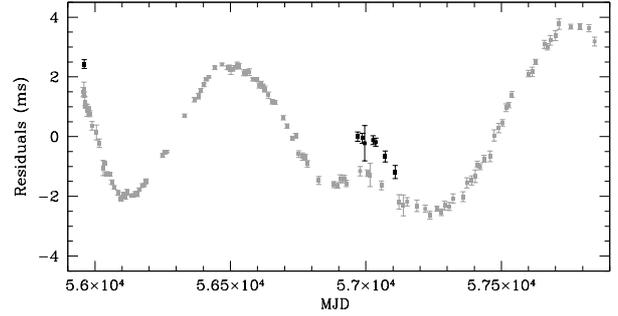}
\caption{Timing residuals for PSR J1757$-$1500. The black points are those
  to which a phase jump needs to be applied to re-align them
  with the rest (grey points) and summing which, the bottom profile
  in Fig \ref{fig:singlepeaked} was created. The last three data
  points are not shown in the plot, to keep the vertical axis smaller
  and the phase jump visible.}
\label{fig:J1757}
\end{figure}

When timing PSR J1757$-$1500, a faint pulsar with a 0.18 s spin
period, we noticed that a small number of timing points appeared
offset with respect to the rest (see Fig. \ref{fig:J1757}). Fitting a
constant phase jump (from MJDs 56838 to 57238, around the
epoch where most of the offset points are, to avoid the effects of
timing noise; see \ref{sec:noise}), we obtain a value of 0.00179(6) s,
compatible with the separation of $\sim$ 0.01 in pulse phase between
the two components of the pulse profile. 

\begin{figure}
\begin{center}
\includegraphics[width=8.5cm,clip]{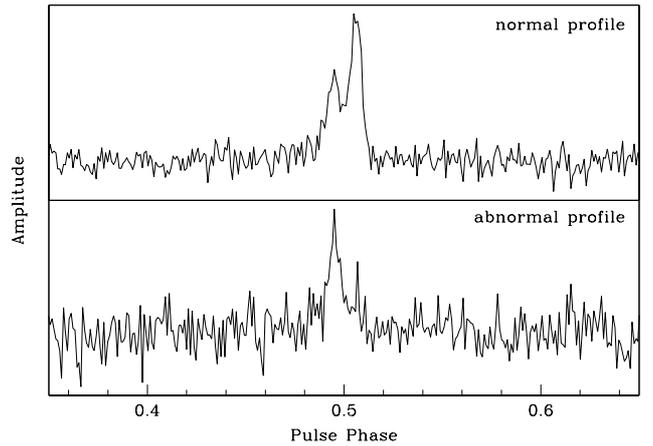}
\end{center}
\caption{Comparison between the standard `normal' pulse profile for
  PSR J1757$-$1500 (top panel) and the profile obtained summing in
  phase only the offset black points in figure \ref{fig:J1757}. The
  profiles are zoomed in to the phase range 0.38--0.62.}
\label{fig:singlepeaked}
\end{figure}

Upon closer inspection, indeed, the (5-minute long) observations to
which the phase jump had to be applied appear, when the noise in the
profile is low enough to make a distinction, to be single peaked. An
integrated pulse profile obtained by summing all the offset
observations can be seen in the bottom panel of Figure
\ref{fig:singlepeaked}.
This behaviour suggests that PSR J1757$-$1500 is sporadically exhibiting mode
changing. Profile changes are observed in pulsars in a number of ways,
from short term mode-changing \citep{bac70a} to long-term changes as
e.g. in intermittent pulsars \citep{klo+06} or even in regular
pulsars \citep{lhk+10}. 

Because of the low S/N of the profiles of the individual observations
(on average 8, over an average integration time of 15 minutes),
assessing more specific details for the profile changes is,
unfortunately, not easily possible: we do not have enough
statistics. This prevents us from obtaining information on possible
changes occurring on time scales shorter than the individual
observations themselves. This, in turn, permits us to only place a
lower limit on the frequency of the occurrence of the single-peaked
mode, present in 8 observations out of 128, all but one of which
concentrated within 138 days. This apparent sporadicity is at odds
with what observed in other moding pulsars, in which the profile
changes occur on shorter times scales and with a somewhat more regluar
cadence. These changes may be more like those seen e.g. \citet{djw+18}
for PSR J1119$-$6127, or by \citet{krj+11} in PSR J0738$-$4042.

The small number and short duration of the observations showing offset
ToAs, hence having a detectable profile variation, makes also
impossible to check if mode changes are accompanied by spin-down
changes, as observed e.g. for intermittent pulsars
\citep{klo+06,lsf+17} and and as theorised to explain nulling and
mode-changing phenomena \citep{tim10}.

\subsection{Pulsars with timing instabilities}

\subsubsection{A glitch from PSR J1635$-$2616}
\label{sec:glitch}

Glitches are still poorly understood instabilities in the otherwise
regular spin evolution of pulsars: they manifest themselves as sudden
changes of the pulsar rotation, usually followed by a slow recovery
towards the pre-glitch values.

PSR J1635$-$2616 showed a glitch around MJD 56458(6), with a frequency
variation $\delta_{\nu} = 1.35(5) \times 10^{-9}$ and a barely
significant variation of the spin-down rate $\delta_{\dot{\nu}} = 5(2)
\times 10^{-18}$ (the errors in parentheses are quoted at
1-$\sigma$). With a characteristic age of 1.9 Myr and a single glitch
in approximately 6.5 years, the glitch rate is typical for this type
of pulsar, and while it is a small glitch, it is not unusual for the
observed period derivative \citep{elsk11}.

\begin{figure}
\includegraphics[width=8.5cm]{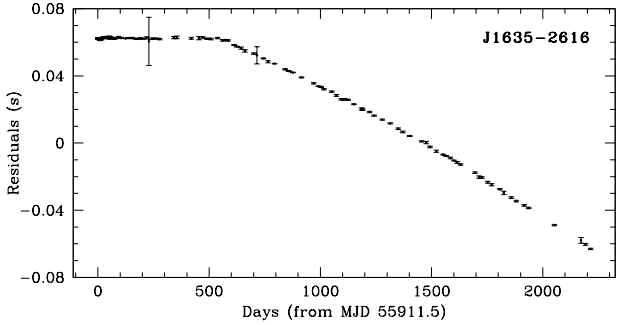}
\caption{Timing residuals for PSR J1635$-$2616 relative to the timing
  solution before the glitch.}
\label{fig:glitch}
\end{figure}

Figure \ref{fig:glitch} shows the timing residuals for PSR
J1635$-$2616 relative to the timing solution before the glitch.

\subsubsection{Timing noise}
\label{sec:noise}

\begin{figure}
\includegraphics[width=8.5cm]{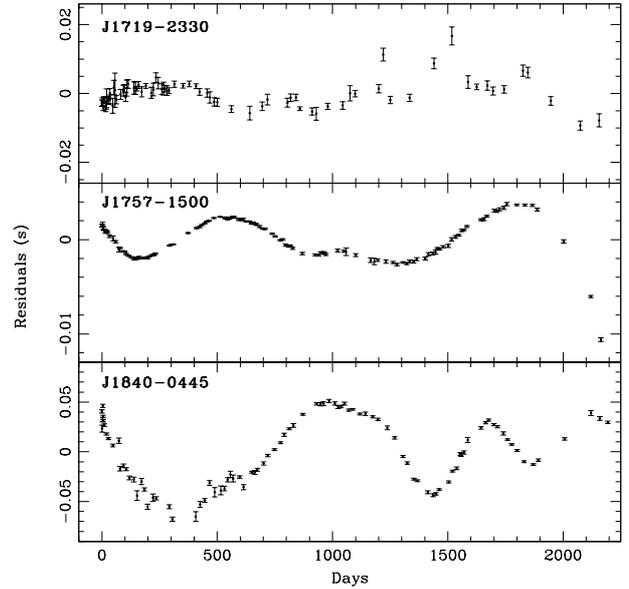}
\caption{Timing residuals for the three noisy pulsars of our sample,
  after fitting for astrometric and spin parameters up to
  $\ddot{\nu}$. Day 0 is the beginning of the data span as reported in
  Tab. \ref{tab:timing}, column 9.}
\label{fig:noise}
\end{figure}

Among the pulsars that we have timed for multiple years, four
(J1719$-$2330, J1757$-$1500, J1840$-$0445 and J1902$-$1036) show
unmodelled long-term instabilities in their rotation (timing noise;
see e.g. \citealt{hlk06}). PSR J1902$-$1036 shows only a
relatively small amount of timing noise so that its timing residuals are
easily ``whitened'' by adding to the fit the spin frequency second
derivative $\ddot{\nu}$ only. To flatten the residuals of other three
pulsars, instead, we had to fit multiple extra frequency
derivatives. Figure \ref{fig:noise} shows the timing residuals for
these three pulsars after the cubic term only ($\ddot{\nu}$) has been
removed.

Table \ref{tab:noise} reports the fitted value of $\ddot{\nu}$, the
pre- and post-whitening rms of the residuals, the number of spin
frequency derivatives used to flatten the timing residuals, and the
values of the stability parameter
$\Delta_8=\log_{10}[t^3|{\ddot{\nu}}|\times 1/(6\nu)]$ \citep{antt94},
(arbitrarily) measured over data spans of 10$^8$ s.

Our data for the pulsars affected by timing noise cover close to twice
$10^8$ s; we report the average of the $\Delta_8$ parameters obtained
for the first and the last 10$^8$ s of data. The $\Delta_8$ values for
our three pulsars are in line with what is observed for other pulsars
affected by timing noise; in particular our small sample appears to
follow the relation between timing noise and spin down rate first
found by \citet{antt94} and refined by \citet{hlk10}.

\begin{table}
\caption{For each pulsar affected by timing noise, the table lists the
  frequency second derivative (at the epochs reported in Tab
  \ref{tab:timing}), the rms of the timing residual obtained by
  fitting up to $\ddot{\nu}$, the rms of the whitened residuals, the
  number of spin frequency derivatives fitted for to whiten the timing
  residuals, and the average stability parameter. Number in
  parentheses are {\tt{tempo2}} 2-sigma errors on the last quoted
  digit(s). Note that, over 10$^8$ s, PSR J1902$-$1036 does not show
  timing noise, hence no $\Delta_8$ is reported.}
\label{tab:noise}
\begin{center}
\begin{tabular}{cccccc}
\hline
PSR & $\ddot{\nu}$ & rms$_{\rm{pre}}$ & rms$_{\rm{post}}$  & \#$\nu$ & $\Delta_8$ \\
    & (10$^{-25}$ s$^{-3}$) & (ms) & (ms) & & \\
\hline
J1719$-$2330 & $-$4.3(10) &  3.15 & 1.27 &  4 & $-$1.41 \\
J1757$-$1500 & $-$7.2(12) &  1.82 & 0.13 & 11 & $-$1.28 \\
J1840$-$0445 & $-$137(7) & 28.31\phantom{0} & 3.08 & 12 & $-$0.9 \\
J1902$-$1036 & $-$0.54(11) & 0.99 & -- & 2  & -- \\
\hline
\end{tabular}
\end{center}
\end{table}

A magnetospheric origin for timing noise has been proposed by
\citet{lhk+10} who showed that, in several cases, the observed spin
irregularity can derive from abrupt changes in the spin-down rate,
typically between two values, which are sometimes linked to pulse shape
changes. A future study, using longer data-spans, will investigate if
this is the case also for these noisy HTRU-med pulsars.

\section{Comparison between HTRU-med and other surveys} 
\label{sec:cfr}

The High Time Resolution Universe Survey has been one of the most
succesful pulsar search experiments of the recent years, especially in
terms of millisecond pulsar discoveries. Its intermediate latitude
section, covering the Galactic latitude range $-15^{\circ}$ to
$+15^{\circ}$, in particular, has discovered 113 new pulsars to date,
29 of which are (fully and mildly) recycled pulsars.

Based on the ATNF pulsar catalogue, 1135 Galactic radio pulsars were
already known in the survey area, 42 of which were recycled
objects. The ratio of recycled over ordinary PSRs in HTRU-med is more
than 7 times higher than that of the previously known sample. Even the
ratio for the previous Parkes experiment covering roughly the same
area, the Swinburne Intermediate Latitude pulsar survey
\citep{ebvb01}, mainly aimed at finding MSPs, was 3 times smaller than
that of HTRU-med.

This confirms that our survey digital signal processor (BPSR), with a
$8\times$ better frequency resolution and a $> 2\times$ better timing
resolution than the major previous Parkes pulsar surveys, was very
well suited to find the rarer, faster MSPs, which often prove to be
interesting and extreme objects. The only other pulsar survey,
although covering a different area of the sky, that can compare to
HTRU-med in terms of MSP discovery rate, is the Arecibo PALFA survey
\citep{cfl+06}, with 17\% of recycled pulsars in 188
discoveries{\footnote{http://www.naic.edu/~palfa/newpulsars/}}. This
similarity is not surprising, since PALFA works at the same frequency
and has almost equal timing and spectral resolution to HTRU-med.

Another interesting number to consider, is the dispersion measure over
spin period ratio DM/$P$, a parameter that can be used to estimate the
depth of a MSP search. DM/$P$ of HTRU-med recycled pulsars, is, on
average, greater than that of previous experiments covering the same
regions of the sky, confirming the effectiveness of our observing
strategy. The mean value of DM/$P$ for HTRU-med discoveries is, indeed,
90\% larger than that of the previously known recycled pulsars in the
area, and the median value is 120\% larger.

\begin{figure}
\includegraphics[width=4.2cm]{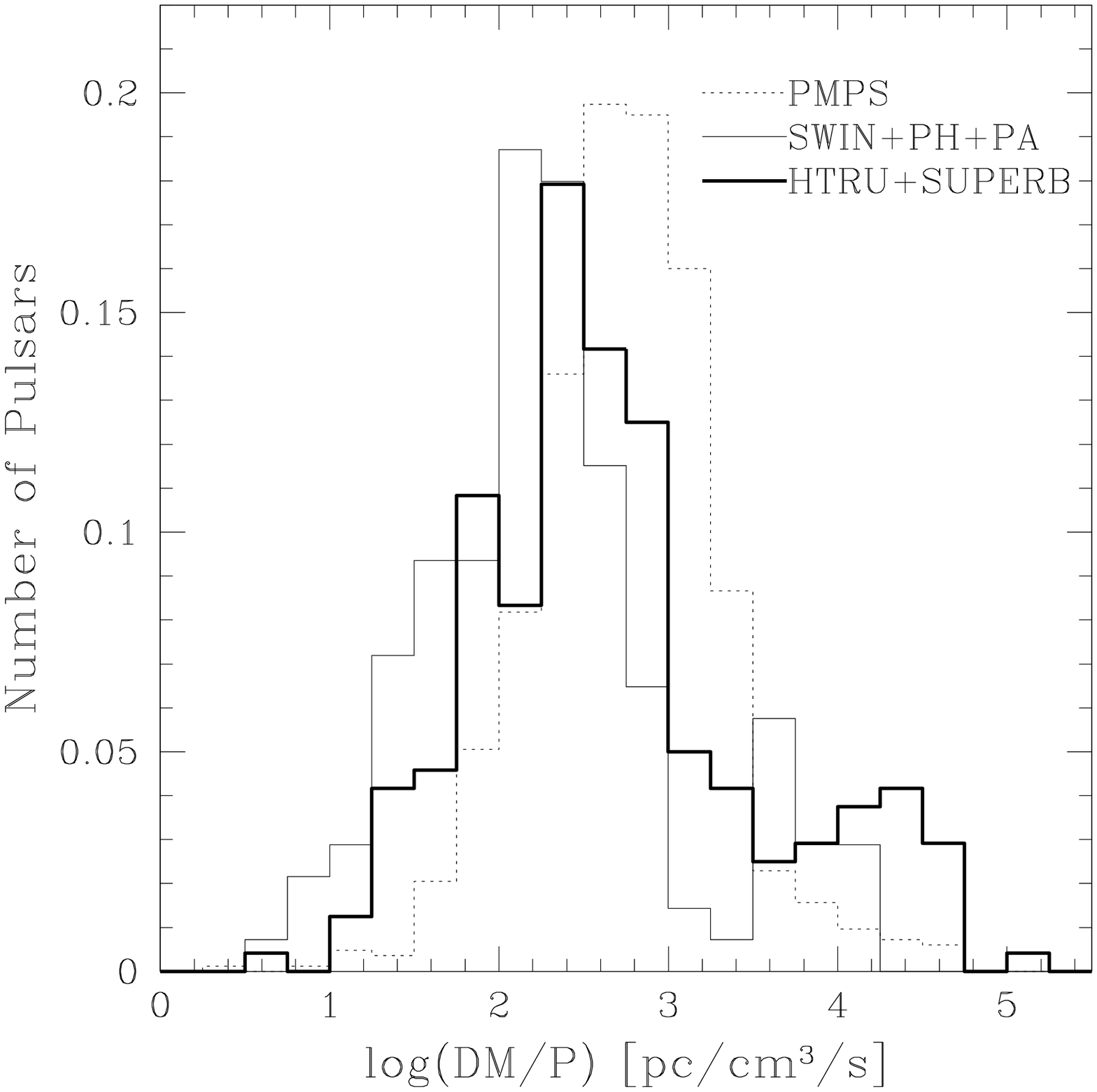}
\includegraphics[width=4.2cm]{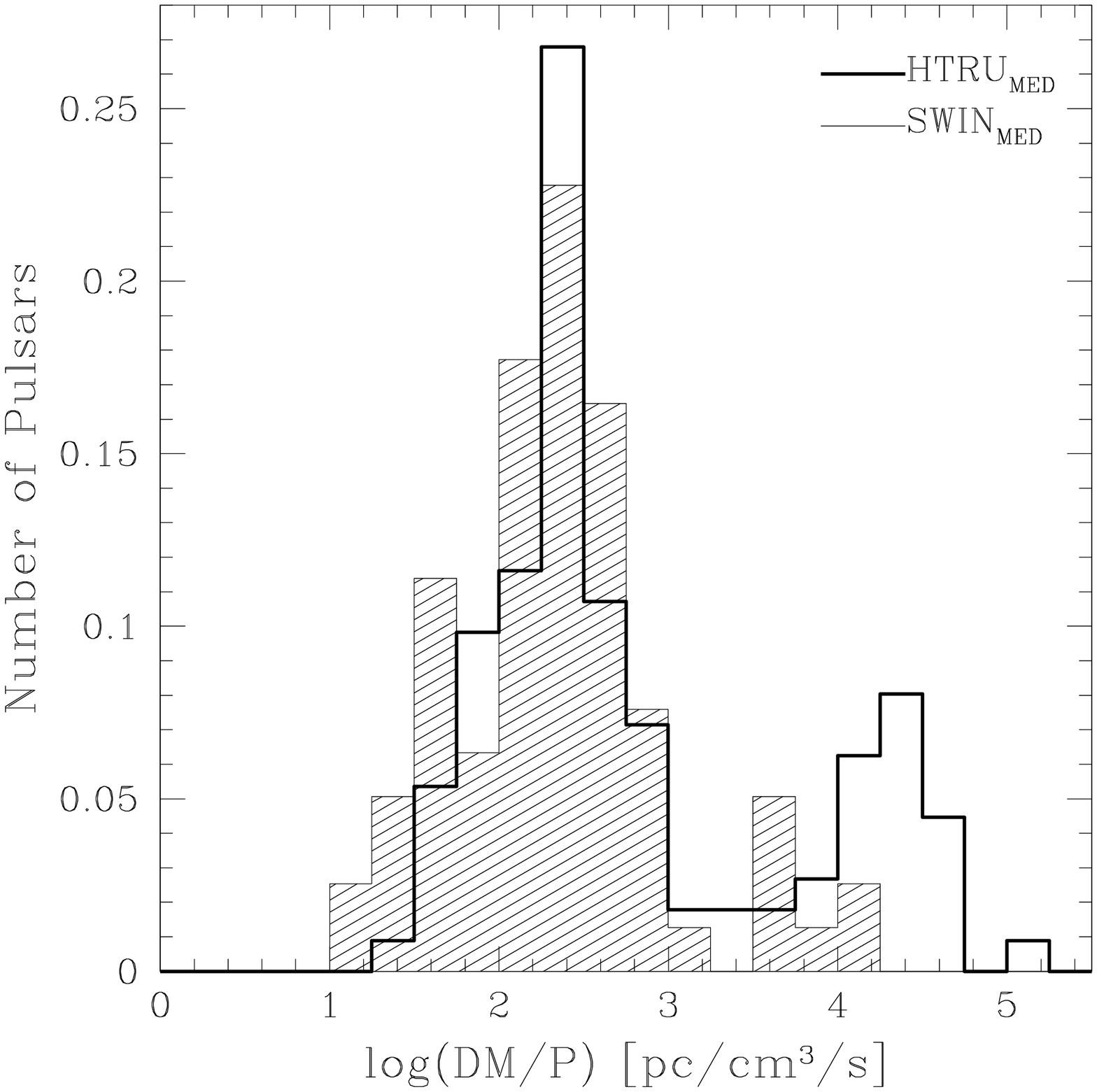}
\includegraphics[width=4.2cm]{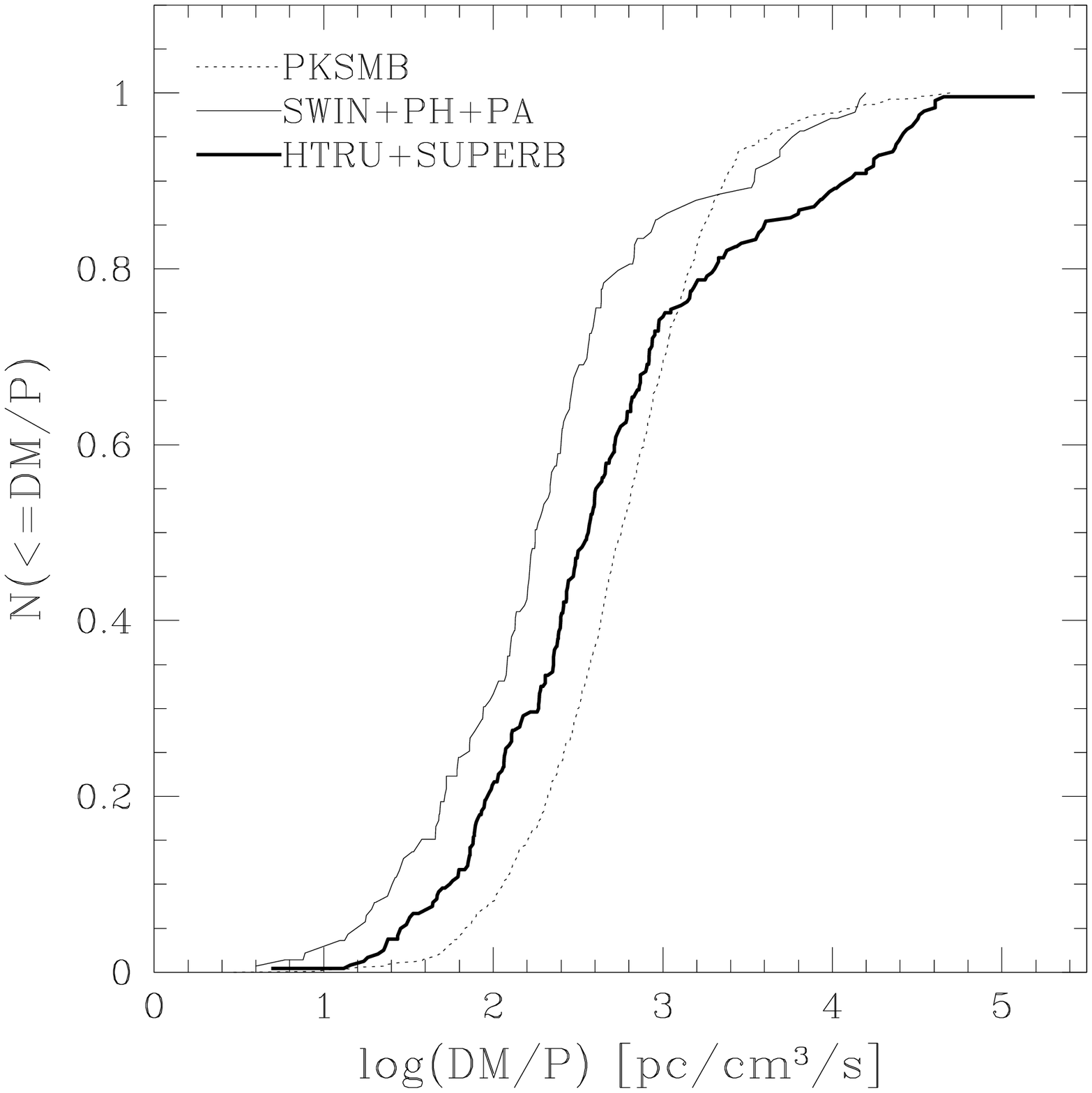}
\includegraphics[width=4.2cm]{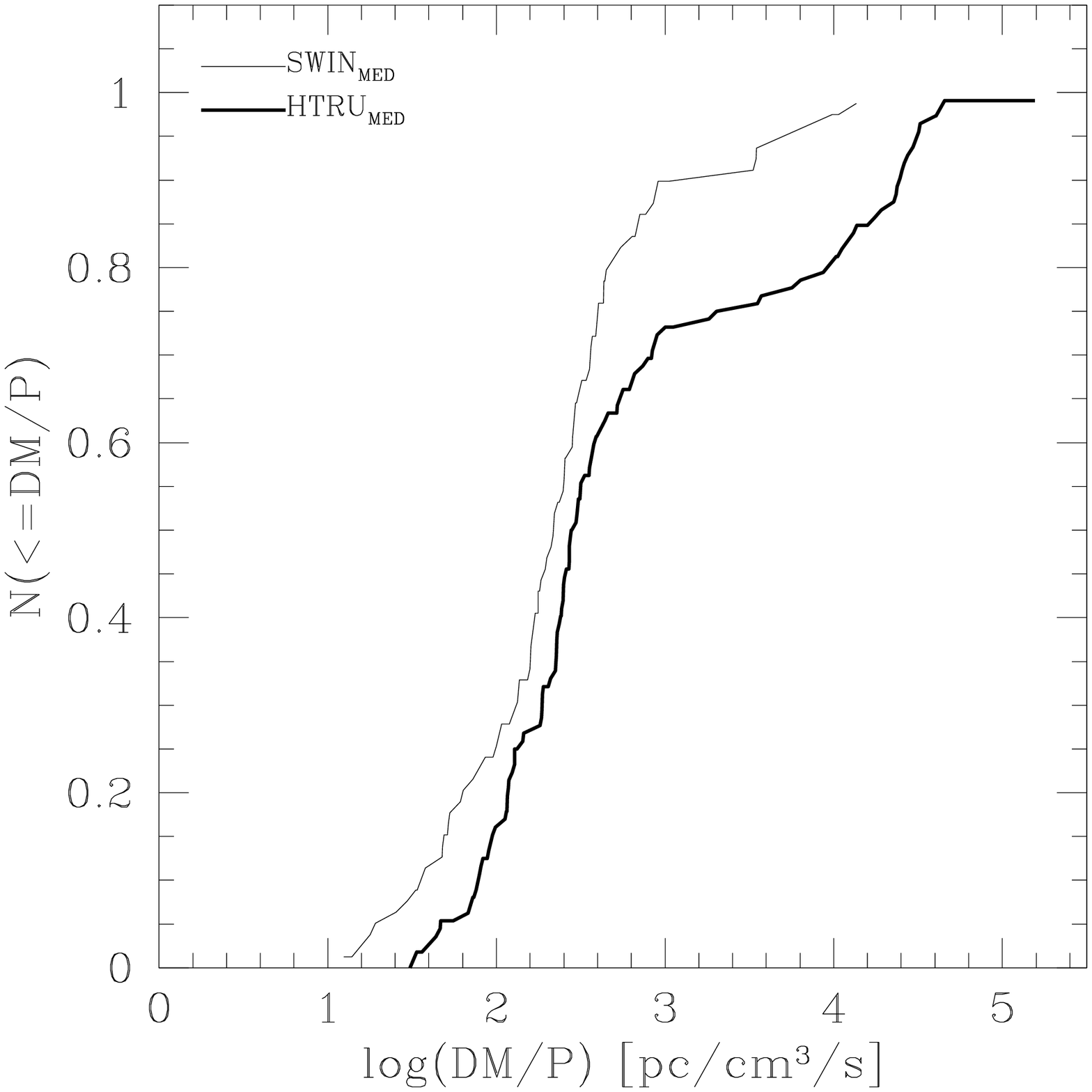}
\caption{Normalised histograms (top) and cumulative distributions
  (bottom) of DM/$P$ for Parkes survey discoveries. Left: the dotted
  curve is for the galactic plane multibeam survey PMPS
  \citep{mlc+01}, with a 3-MHz frequency resolution and a 250 $\mu$s
  time resolution; the solid thin curve is for surveys with the same
  frequency resolution, but 2x finer time resolution (SWIN,
  \citealt{ebvb01,jbo+09}; PH, \citealt{bjd+06} and PA,
  \citealt{bkl+13}); the solid thick curve is for the most recent
  surveys using the BPSR digital backend, with 0.39-MHz frequency
  resolution and 64 $\mu$s time resolution (HTRU, \citealt{kjv+10} and
  SUPERB, \citealt{kbj+18}). Right: comparison of
  intermediate-latitude Parkes surveys only. The shaded histogram is
  for the SWIN surveys, the thick solid line for HTRU-med.}
\label{fig:dmp}
\end{figure}

Figure \ref{fig:dmp} shows, on the left, a comparison of the DM/$P$
histograms for Parkes surveys of incresing time and frequency
resolution: from the galactic plane multibeam-survey (PMPS;
\citealt{mlc+01}), that used an analogue filterbank with 3-MHz
channels and a sampling time of 250 $\mu$s, to the high and
intermediate-latitude surveys (PH and SWIN;
\citealt{ebvb01,bjd+06,jbo+09}), that, together with PMPS extension on
the Perseus Arm (PA; \citealt{bkl+13}), used the same backend, but a
125 $\mu$ sempling time, to the HTRU and SUPERB \citep{kbj+18}
surveys, using the BPSR digital backend with 0.39-MHz channels and a
64 $\mu$s sampling time. While PMPS, covering the denser plane of the
Galaxy, has a larger fraction of non-recycled, high DM pulsars, the
highest DM/$P$ objects, with small spin periods and large DMs, increase
with the finer time and frequency resolution of subsequent
experiments. The improvement (especially for recycled pulsar) is
clearer and unbiased if we compare (Fig. \ref{fig:dmp}, right) the
intermediate latitude part of HTRU only, with SWIN$_{\rm{MED}}$,
covering the same area in the sky (hence regions with the same free
electron content). We note that we chose to show only discoveries, and
not also redetections, because the latter can be biased by the prior
knowledge of the existence of a pulsar with a specific $P$ and DM: even
with a low S/N and a broadened profile, a previoulsy known pulsar can
be safely recognised and listed as redetection, while, if it were a
new source, it might not even be considered as a candidate worth of
reobservation.

The HTRU-med MSP J1804$-$2858, with a $P$ of 1.49 ms and a DM of 232
pc/cm$^3$\citep{mbc+19}, is the second fastest pulsar discovered in
the Galactic field, and has the second highest DM/P value of all kown
pulsars (the record being held by the Globular Cluster MSP
J1748$-$2446ad; \citealt{hrs+06}). In PMPS the pulse of such an
extreme object would have been completely whashed away, the
DM-smearing being larger than the spin period, in a 3-MHz channel.
Future surveys, such as the upcoming experiments with the Square
Kilometre Array presursor MeerKAT \citep{jon09}, will be sensitive to
even faster and/or further away objects: a pulsar with the same
DM-smearing that J1804$-$2858 has in HTRU, for instance, can be found,
with MeerKAT, at DMs 50\% larger (hence, roughly, 50\% more distant,
in a similarly dense environment); on the other hand, a pulsar with
the same DM as J1804$-$2858 will have the same observed duty-cycle as
the one J1804$-$2858 has in HTRU, even with a 50\% shorter spin
period, thanks to the improved frequency resolution of MeerKAT
surveys, which will use 4096 channels over a 856 MHz bandwidth.

\section{Conclusions}
\label{sec:sum}

In this paper we have presented the discovery and timing of six more
pulsars found in the intermediate latitude part of the High Time
Resolution Universe Survey, and a coherent timing solution for a
further 19 HTRU-med pulsars. With these discoveries, the last coming
from our standard data analysis pipeline, the total yield for HTRU-med
is now 113 new pulsars, 29 of which - a record 25\% of the total - are
recycled objects. HTRU-med has hence proven to be very effective at
finding new MSPs and in particular, thanks to its time and frequency
resolution, in finding objects covering previously poorly populated
parameter spaces (e.g. high DMs and low P).

Among HTRU-med discoveries we can list many notable pulsars: from the
binary MSP J1801$+$3212 \citep{nbb+14}, potentially useful to test the
Strong Equivalence Principle, to the `diamond planet' pulsar
J1719$-$1438 \citep{bbb+11a}, shedding light on the formation
mechanism of ultra-compact systems. The long period, mildly eccentric
binary pulsar J1837$-$0822 (this work) is now the third of a small
class of peculiar objects whose companion has likely lost mass without
spinning the NS up. Its discovery adds a piece to the puzzle of the
evolution of these systems. Also interesting for the study of binary
evolution are the eclipsing Redbacks and Black Widow pulsars - one of
each was found in HTRU-med \citep{btb+15,bbb+11}. Their final fate is
believed to lead to the formation of isolated MSPs. Two fully recycled
such objects were found in the survey area (this work and
\citealt{bbb+13}). Analysing eclipsing `spider' pulsar characteristics
and comparing them with those of fully recycled isolated MSPs, we
found contrasting evidences: on one hand, we showed that the spin
period of isolated MSPs is significantly longer than that of
spiders. This, simply taking into account the dipolar spin-down, would
corroborate the idea that isolated MSPs may represent a later
evolutionary stage of the faster spinning eclipsing binaries. On the
other hand, however, isolated pulsars are found, on average, further
away from the Galactic plane than Redbacks and Black Widows. If the
two shared a common evolutionary path, they should have received a
similar natal kick, which would be reflected in a similar spatial
distribution.

The positive outcome of HTRU-med, finally, is further highlighted by
the fact that six of our MSPs have been already included in Pulsar
Timing Array projects, which was one of the main goals of the
intermediate part of our experiment \citep{kjv+10}.

The next step forward to find more distant, fainter, higher DM
recycled pulsars will be achieved by the new surveys planned with the
South African SKA precursor MeerKAT. Thanks to its superb sensitivity,
its larger bandwidth and the possibility for finer time and frequency
resolution, MeerKAT has the potential to discover many new peculiar
and extreme objects\citep{sk16}.

\section*{Acknowledgements}
The Parkes radio telescope is part of the Australia Telescope which is
funded by the Commonwealth of Australia for operation as a National
Facility managed by CSIRO. OzGrav is funded by the Australian research
Council under grant CE170100004. RPE gratefully acknowledges support
from ERC Synergy Grant ``BlackHoleCam'' Grant Agreement Number 610058
(PIs: H. Falcke, M. Kramer, L. Rezzolla). Pulsar research at the
Jodrell Bank Centre for Astrophysics and the observations using the
Lovell Telescope are supported by a consolidated grant from the STFC
in the UK. We thank Cees Bassa for his work on the ROACH backend used
for the Lovell observations.








\bsp	
\label{lastpage}
\end{document}